\documentclass[usenatbib,usegraphicx]{mn2e}

\usepackage{psfig}
\usepackage{color}

\def\apj{ApJ}
\def\apjl{ApJL}
\def\apjs{ApJS}
\def\aap{A\&A}

\def\mnras{MNRAS}
\def\aj{AJ}
\def\araa{ARA\&A}
\def\nat{Nature}
\def\pasj{PASJ}
\def\jcap{JCAP}

\def\Al26{\textrm{$^{26}$Al}}
\def\Fe60{\textrm{$^{60}$Fe}}
\def\cm{\textrm{cm}}

\def\micron{\mu\textrm{m}}

\def\kpc{\textrm{kpc}}
\def\pc{\textrm{pc}}
\def\Mpc{\textrm{Mpc}}

\def\Kelv{\textrm{K}}

\def\gcm2{\textrm{g}~\textrm{cm}^{-2}}
\def\kms{\textrm{km}~\textrm{s}^{-1}}
\def\phcm2s1{\textrm{photons}~\textrm{cm}^{-2}~\textrm{s}^{-1}}

\def\eV{\textrm{eV}}

\def\MeV{\textrm{MeV}}

\def\yr{\textrm{yr}}
\def\Myr{\textrm{Myr}}
\def\Gyr{\textrm{Gyr}}
\def\kyr{\textrm{kyr}}

\def\Msun{\textrm{M}_{\sun}}
\def\Lsun{\textrm{L}_{\sun}}
\newcommand{\mean}[1]{\ensuremath{\langle #1 \rangle}}
\def\bfnop{}

\title[Starbursts are Radioactive]{Starbursts and High-Redshift Galaxies are Radioactive: High Abundances of $^{26}$A\lowercase{l} and Other Short Lived Radionuclides}
\author[Lacki]{Brian C. Lacki$^{1,2}$\\$^1$Jansky Fellow of the National Radio Astronomy Observatory\\$^2$Institute for Advanced Study, Einstein Drive, Princeton, NJ 08540, USA, brianlacki@ias.edu}

\begin{document}
\maketitle

\begin{abstract}
Short lived radionuclides (SLRs) like \Al26 are synthesized by massive stars and are a byproduct of star formation.  The abundances of SLRs in the gas of a star-forming galaxy {\bfnop are} inversely proportional to {\bfnop the} gas consumption time.  The rapid evolution of specific star formation rate (SSFR) of normal galaxies implies they had mean SLR abundances $\sim 3$ -- $10$ times higher at $z = 2$.  During the epoch of Solar system formation, the background SLR abundances of the Galaxy were up to twice as high as at present, if SLR yields from massive stars do not depend on metallicity.  If SLRs are homogenized in the gas of galaxies, the high SSFRs of normal galaxies can partly explain the elevated abundance of SLRs like \Fe60 and \Al26 in the early Solar system.  Starburst galaxies have much higher SSFRs still, and have enormous mean abundances of \Al26 (\Al26/$^{27}$Al $\approx 10^{-3}$ for Solar metallicity gas).  The main uncertainty is whether the SLRs are mixed with the star-forming molecular gas: they could be trapped in hot gas and decay before entering the colder phases, or be blown out by starburst winds.  I consider how variability in star-formation rate affects the SLR abundances, and I discuss how SLR transport may differ in these galaxies.  The enhanced \Al26 of starbursts might maintain moderate ionization rates ($10^{-18}$ -- $10^{-17}\ \sec^{-1}$), possibly dominating ionization in dense clouds not penetrated by cosmic rays.  Similar ionization rates would be maintained in protoplanetary discs of starbursts, if the SLRs are well-mixed, and the radiogenic heating of planetesimals would likewise be much higher.  In this way, galaxy evolution can affect the geological history of planetary systems.
\end{abstract}

\begin{keywords}
galaxies:starburst --- galaxies:ISM --- galaxies: evolution --- nuclear reactions, nucleosynthesis, abundances --- galaxies: high redshift --- planets and satellites: formation
\end{keywords}

\section{Introduction}
\label{sec:Introduction}

Massive, young stars are sites of nucleosynthesis, not just of the stable nuclides, but radioactive isotopes as well.  Long-lived radioisotopes with decay times of billions of years or longer, such as $^{40}$K, mix into the interstellar medium (ISM), accumulating and provide a very low level source of energetic radiation in all gas \citep{Cameron62,Umebayashi81}.  More unstable isotopes are also synthesized; some have decay times of a few years or less and cannot reach most of the ISM.  In between those two extremes are the short lived radionuclides (SLRs) of the ISM: with $\sim \Myr$ decay times, they can be present in a galaxy's gas but only as long as star formation replenishes their abundances.  The most prominent of the SLRs is \Al26, which is detected in gamma-ray decay lines from star-formation regions throughout the Milky Way \citep{Mahoney84,Diehl06}.  Another SLR that has recently been detected in gamma-ray lines is $^{60}$Fe \citep{Harris05}.

SLRs are not just passive inhabitants of the ISM.  By releasing energetic particles and radiation when they decay, they inject power that can heat or ionize the surrounding matter.  In the Milky Way's molecular clouds, the radioactivity is overwhelmed by that of cosmic rays, which sustain an ionization rate of $\zeta_{\rm CR} \approx 5 \times 10^{-17}\ \sec^{-1}$.  Rapidly star-forming galaxies known as starbursts may have elevated levels of cosmic rays and ionization rates a thousand times or more higher than the Milky Way (\citealt*{Suchkov93}; \citealt{Papadopoulos10-CRDRs}).  However, it is possible that cosmic rays are stopped by the abundant gas in starbursts before they enter the densest molecular gas.  Gamma rays can provide ionization through large columns; but while the gamma-ray ionization rate can reach up to $\zeta_{\gamma} \approx 10^{-16}\ \sec^{-1}$ in the densest starbursts, in most starbursts gamma rays sustain relatively weak ionization \citep{Lacki12-GRDRs}.  SLRs like \Al26 can in principle provide ionization through any column of gas, and if abundant enough, maintain moderate levels of ionization.  The major open question for SLR-induced ionization is how well mixed the SLRs are with the gas of the galaxy, a process which takes anything up to 100 Myr; if the mixing times are longer than a few Myr, the SLRs are not abundant in most star-forming cores \citep{Meyer00,Huss09}.  

Meteorites recording the composition of the primordial Solar system demonstrate that SLRs were present during its formation. Assuming the SLRs were not created in situ by energetic radiation from the Sun \citep{Lee98}, the SLRs provide evidence that the Solar system formed near a star-forming region with young massive stars \citep[e.g.,][]{Adams10}.  In fact, \Al26 was overabundant by a factor $\sim 6$ in the primordial Solar system, with $X (\Al26) \approx 10^{-10}$ ($\Al26 / ^{27}{\rm Al} \approx 5 \times 10^{-5}$), compared to its present day abundances in the Milky Way (e.g., \citealt*{Lee77}; \citealt*{MacPherson95}; \citealt{Diehl06}; \citealt{Huss09}).  Their quick decay time also indicate the Solar system formed quickly, within about a few Myr.  SLRs, particularly \Al26, were a primary source of ionization in the Solar Nebula \citep{Stepinski92,Finocchi97,Umebayashi09}, affecting the conductivity and ultimately accretion rate in the protoplanetary disc.  Moreover, \Al26 and other SLRs may have regulated the early geological evolution of the Solar system by being a major source of heat in early planetesimals, driving their differentiation and rock metamorphism \citep[e.g.,][]{Hutcheon89,Grimm93,Shukolyukov93}.  

The contemporary Milky Way, with a star-formation rate (SFR) of a few stars per year is not a typical environment for most of the star-formation in the history of the Universe, however.  Roughly 5-20\% of star-formation at all times occurred in rapid starbursts mainly driven by galaxy-galaxy interactions and mergers \citep{Rodighiero11,Sargent12}.  Furthermore, most of the star-formation in `normal' galaxies occurred in massive galaxies with a much higher star-formation rate ($\ga 10\ \Msun\ \yr^{-1}$) at redshifts $z$ of 1 and higher, when most star-formation took place \citep[e.g.,][]{Magnelli09}.  These high star-formation rates translate into large masses of SLRs present in these galaxies.  I {\bfnop will} show that \Al26 in these galaxies, if it is well mixed with the gas, can sustain rather high ionization rates in their ISMs.  This has consequences for both star formation and planet formation.

When necessary, I assume a Hubble constant of $H_0 = 72\ \kms\ \Mpc^{-1}$, a matter density of $\Omega_M = 0.25$, and a cosmological constant $\Omega_{\Lambda} = 0.75$ for the cosmology.

\section{The Equilibrium Abundance of SLR\lowercase{s}}
In a one-zone model of a galaxy, which disregards spatial inhomogeneities, the complete equation for the SLR mass $M_{\rm SLR}$ in the ISM is
\begin{equation}
\label{eqn:MeqSLRFull}
\frac{dM_{\rm SLR}}{dt} = Q_{\rm SLR}(t) - \frac{M_{\rm SLR} (t)}{\tau_{\rm SLR}},
\end{equation}
where $\tau_{\rm SLR}$ is the lifetime of the SLR in the galaxy.  $Q_{\rm SLR} (t)$, the injection rate of the SLR, depends on the past star-formation history:
\begin{equation}
\label{eqn:QSLR}
Q_{\rm SLR} (t) = \int_{-\infty}^t Y_{\rm SLR}(t - t^{\prime}) \times {\rm SFR}(t^{\prime}) dt^{\prime}. 
\end{equation}
For a coeval stellar population of age $t$, the yield $Y_{\rm SLR}(t)$ is the mass ejection rate of the SLR into the interstellar medium per unit stellar mass \citep{Cervino00}.

If there are no big fluctuations in the star-formation rate over the past few Myr, then the SLR abundance approaches a steady-state.  The equilibrium mass of a SLR in a galaxy is proportional to its star-formation (or supernova) rate averaged over the previous few Myr.  We can then {\bfnop parametrize} the injection of SLRs in the ISM by a yield $\Upsilon_{\rm SLR}$ per supernova:
\begin{equation}
\Upsilon_{\rm SLR} = \varepsilon \int_0^{\infty} Y(t^{\prime\prime}) dt^{\prime\prime},
\end{equation}
regardless of whether SNe are actually the source of SLRs.  The $\varepsilon$ factor is the ratio of the supernova rate $\Gamma_{\rm SN}$ and star-formation rate.  Then the equilibrium SLR mass is given by \citep[e.g.,][]{Diehl06}
\begin{equation}
\label{eqn:MeqSLR}
M_{\rm SLR}^{\rm eq} = \Gamma_{\rm SN} \Upsilon_{\rm SLR} \tau_{\rm SLR}.
\end{equation}
The supernova rate is proportional to the star-formation rate, so $\Gamma_{\rm SN} = \varepsilon {\rm SFR}$.  The abundance of an SLR is given by $X_{\rm SLR} = M_{\rm SLR}^{\rm eq} m_H / (M_{\rm gas} m_{\rm SLR})$, where $m_{\rm SLR}$ is the mass of one atom of the SLR and $M_{\rm gas}$ is the gas mass in the galaxy.  Therefore the abundance of an SLR is
\begin{equation}
X_{\rm SLR} = \varepsilon \frac{\rm SFR}{M_{\rm gas}} \frac{\Upsilon_{\rm SLR} \tau_{\rm SLR} m_H}{m_{\rm SLR}}
\end{equation}
The quantity $M_{\rm gas} / {\rm SFR} = \tau_{\rm gas}$ is the gas consumption time.  Note that it is related to the specific star formation rate, ${\rm SSFR} = M_{\star} / {\rm SFR}$, as $\tau_{\rm gas} = f_{\rm gas} / ((1 - f_{\rm gas}) {\rm SSFR})$, where $M_{\star}$ is the stellar mass and $f_{\rm gas} = M_{\rm gas} / (M_{\rm gas} + M_{\star})$ is the gas fraction.  Therefore, we can express the equilibrium mass of the SLR in a galaxy as
\begin{equation}
X_{\rm SLR} = \frac{\varepsilon \Upsilon_{\rm SLR} \tau_{\rm SLR} m_H}{\tau_{\rm gas} m_{\rm SLR}} = \varepsilon \frac{1 - f_{\rm gas}}{f_{\rm gas}} {\rm SSFR} \frac{\Upsilon_{\rm SLR} \tau_{\rm SLR} m_H}{m_{\rm SLR}}
\end{equation}
Finally, the ratio of SLR abundance in a galaxy to that in the Milky Way is
\begin{eqnarray}
\nonumber \frac{X_{\rm SLR}}{X_{\rm SLR}^{\rm MW}} & = & \frac{\tau_{\rm gas}^{\rm MW}}{\tau_{\rm gas}} \frac{\tau_{\rm SLR}}{\tau_{\rm SLR}^{\rm MW}}\\
& = & \frac{1 - f_{\rm gas}}{1 - f_{\rm gas}^{\rm MW}} \frac{f_{\rm gas}^{\rm MW}}{f_{\rm gas}} \frac{\rm SSFR}{\rm SSFR^{\rm MW}} \frac{\tau}{\tau^{\rm MW}},
\end{eqnarray}
with a MW superscript referring to values in the present day Milky Way.  Thus, galaxies with short gas consumption times (and generally those with high SSFRs) should have high abundances of SLRs.  The reason is that in such galaxies, more of the gas is converted into stars and SLRs within the residence time of an SLR.

The greatest uncertainty in these abundances is the residence time $\tau_{\rm SLR}$.  In the Milky Way, these times are just the radioactive decay times, defined here as the e-folding time.  In starburst galaxies, however, much of the volume is occupied by a hot, low density gas which forms into a galactic wind with characteristic speeds $v$ of several hundred {\bfnop kilometres} per second (e.g., \citealt{Chevalier85}; \citealt*{Heckman90}; \citealt{Strickland09}).  If massive stars emit SLRs at random locations in the starburst, most of them will dump their SLRs into the wind phase of the ISM.  The wind-crossing time is $\tau_{\rm wind} = 330\ \kyr\ (h / 100\ \pc) (v / 300\ \kms)^{-1}$, where $h$ is the gas scale-height.  The equilibrium time in starburst galaxies is then $\tau = [\tau_{\rm decay}^{-1} + \tau_{\rm wind}^{-1}]^{-1}$.  Furthermore, the SLRs ejected into the wind may never mix with the molecular gas, so the fraction of SLRs injected into the molecular medium may be $\ll 1$ (I discuss this issue further in section~\ref{sec:Mixing}).  However, very massive stars are found close to their birth environments where there is a lot of molecular gas to enrich, and these may be the source of \Al26, as supported by the correlation of the 1.809 MeV \Al26 decay line emission and free-free emission from massive young stars \citep{Knoedlseder99}.

Turning to the specific example of \Al26, I note that the yield of \Al26 is thought to be $\Upsilon_{\rm Al-26} \approx 1.4 \times 10^{-4}\ \Msun$ per supernova \citep{Diehl06}.  For a Salpeter initial mass function from $0.1 - 100\ \Msun$, the supernova rate is $\Gamma_{\rm SN} = 0.0064\ \yr^{-1} ({\rm SFR} / \Msun\ \yr^{-1})$, or $\Gamma_{\rm SN} = 0.11\ \yr^{-1} (L_{\rm TIR} / 10^{11}\ \Lsun)$ in terms of the total infrared ($8 - 1000\ \micron$) luminosity $L_{\rm TIR}$ of starbursts (\citealt{Kennicutt98}; \citealt*{Thompson07}).  If I suppose all of the \Al26 is retained by the molecular gas, so that the residence time is the \Al26 decay time of 1.04 Myr, then the equilibrium abundance of \Al26 in a galaxy is just
\begin{equation}
\label{eqn:XAl26Numer}
X (\Al26) = 3.4 \times 10^{-11} \left(\frac{\tau_{\rm gas}}{\Gyr}\right)^{-1} = 1.7 \times 10^{-9} \left(\frac{\tau_{\rm gas}}{20\ \Myr}\right)^{-1}
\end{equation}

\subsection{High-Redshift Normal Galaxies}
\label{sec:MSGalaxies}

The star-formation rates and stellar masses of normal star-forming galaxies lie on a `main sequence' with a characteristic SSFR that varies weakly, if at all, with stellar mass \citep[e.g.,][]{Brinchmann04}.  However, the characteristic SSFR evolves rapidly with redshift \citep[e.g.,][]{Daddi07,Noeske07,Karim11}, with ${\rm SSFR} \propto (1 + z)^{2.8}$ out to $z \approx 2.5$ -- a rise of factor $\sim 30$ \citep{Sargent12}.  At $z \ga 2.5$, the SSFR of the main sequence then seems to remain constant \citep{Gonzalez10}.  

Countering this rise in the SSFR, the gas fractions of normal galaxies at high $z$ were also higher: the high equilibrium masses of SLRs are diluted to some extent by higher gas masses.  \citet{Hopkins10} provide a convenient equation, motivated by the Schmidt law \citep{Kennicutt98}, to describe the evolution of gas fraction:
\begin{equation}
\label{eqn:fGas}
f_{\rm gas} (z) = f_0 [1 - (t_L (z) / t_0) (1 - f_0^{3/2})]^{-2/3},
\end{equation}
assuming a gas fraction $f_0$ at $z = 0$, with a look back time of $t_L (z) = \int_0^z dz^{\prime} / [H_0 (1 + z^{\prime}) \sqrt{\Omega_{\Lambda} + \Omega_M (1 + z^{\prime})^3}]$ and a current cosmic age of $t_0$ \citep[see also][]{Hopkins09}.  Since the gas fractions of normal galaxies at present are small, the evolution at low redshifts can be approximated as $f_{\rm gas} (z) = f_0 [1 - (t_L (z) / t_0)]^{-2/3}$.  

\begin{figure}
\centerline{\includegraphics[width=8cm]{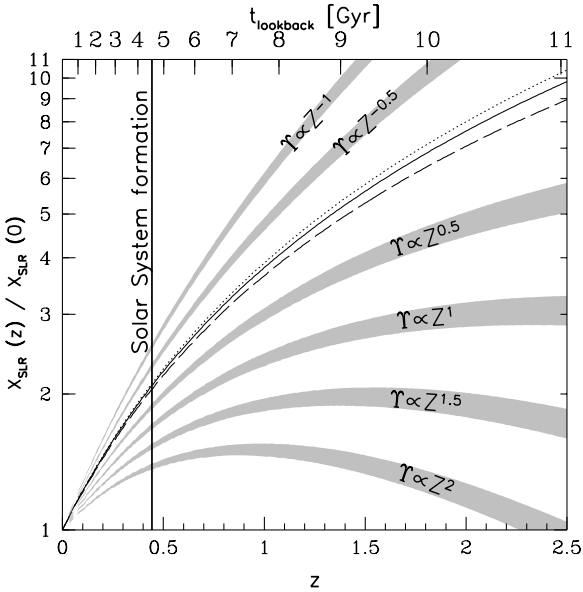}}
\caption{Plot of the SLR abundance enhancements in normal galaxies lying on the `main sequence', for a gas fraction evolution described by equation~\ref{eqn:fGas}.  The rapid evolution of SSFRs leads to big enhancements of SLRs at high $z$.  Even during the epoch of Solar system formation, the mean SLR abundance was twice the present value.  The different lines are for different $f_{\rm gas}$ at $z = 0$, assuming SLR yields are independent of metallicity: 0.05 (dotted), 0.1 (solid), 0.2 (dashed).  The shading shows the abundances for $0.05 \le f_{\rm gas} \le 0.2$ when the SLR yield depends on metallicity, assuming a 0.2 dex decrease in metallicity per unit redshift. \label{fig:NormGalaxy}}
\end{figure}

After calculating the mean abundances of SLRs in normal galaxies, I find that the rapid SSFR evolution overwhelms the modest evolution in $f_{\rm gas}$ at high $z$: the SLR abundances of normal galaxies evolves quickly.  These enhancements are plotted in Fig.~\ref{fig:NormGalaxy}.  

Observational studies of high redshift main sequence galaxies indicate a slower evolution in $\tau_{\rm gas}$, resulting from a quicker evolution of $f_{\rm gas}$.  Although equation~\ref{eqn:fGas} implies that $f_{\rm gas}$ was about twice as high ten billion years ago at $z \approx 2$, massive disc galaxies are observed with gas fractions of $\sim 40$ -- $50\%$, which is 3 to 10 times greater than at present \citep[e.g.,][]{Tacconi10,Daddi10}.  According to \citet{Genzel10}, the typical (molecular) gas consumption time at redshifts 1 to 2.5 was $\sim 500\ \Myr$.  In the \citet{Daddi10} sample of BzK galaxies at $z \approx 1.5$, gas consumption times are likewise $\sim 300$ -- $700\ \Myr$.  To compare, the molecular gas consumption times at the present are estimated to be 1.5 to 3 Gyr \citep{Diehl06,Genzel10,Bigiel11,Rahman12}, implying an enhancement of a factor 3 to 6 in SLR abundances at $z \ga 1$.  But note that the BzK galaxies are not the direct ancestors of galaxies like the present Milky Way, which are less massive.  The SSFR, when observed to have any mass dependence, is greater in low mass galaxies at all $z$ \citep{Sargent12}.  This means that lower mass galaxies at all $z$ have shorter $\tau_{\rm gas}$, as indicated by observations of present galaxies \citep{Saintonge11}.  The early Milky Way therefore may have had a gas consumption time smaller than $500\ \Myr$.

So far, I have ignored possible metallicity $Z$ dependencies in the yield $\Upsilon$ of SLRs.  It may be generally expected that star-forming galaxies had lower metallicity in the past, since less of the gas has been processed by stellar nucleosynthesis.  However, observations of the age-metallicity relation of G dwarfs near the Sun reveal that they have nearly the same metallicity at ages approaching 10 Gyr (e.g., \citealt{Twarog80,Haywood06}; \citealt*{Holmberg07}), though the real significance of the lack of a trend remains unclear, since there is a wide scatter in metallicity with age \citep[see the discussion by][]{Prantzos09}.  Observations of external star-forming galaxies find weak evolution at constant stellar mass, with metallicity $Z$ decreasing by $\sim 0.1-0.2$ dex per unit redshift (\citealt*{Lilly03}; \citealt{Kobulnicky04}).

After adopting a metallicity dependence of $Z(z) = Z(0) \times 10^{-0.2 z}$ (0.2 dex decrease per unit redshift), I show in the revised SLR abundances Fig.~\ref{fig:NormGalaxy} assuming that the SLR yield goes as $Z^{-1}$, $Z^{-0.5}$, $Z^{0.5}$, $Z$, $Z^{1.5}$, and $Z^2$.  If the yields are smaller at lower metallicity, the SLR abundances are still elevated at high redshift, though by not as much for metallicity-independent yields.  As an example, the yield of \Al26 in the winds of Wolf-Rayet stars is believed to scale as $\Upsilon \propto Z^{1.5}$ \citep{Palacios05}.  According to \citet{Limongi06}, these stellar winds contribute only a minority of the \Al26 yield, so it is unclear how the \Al26 yield really scales.  \citet{Martin10} considered the \Al26 and \Fe60 yields from stars with half Solar metallicity.  They found that, because reduced metallicity lowers wind losses, more SLRs are produced in supernovae.  This mostly compensates for the reduced wind \Al26 yield, and actually raises the synthesized amount of \Fe60.

\subsection{Starbursts}

\begin{table*}
\begin{minipage}{170mm}
\caption{$^{26}$A\lowercase{l} Abundances and Associated Ionization Rates}
\label{table:Al26Abundances}
\begin{tabular}{lccccccccc}
\hline
Starburst                             & SFR & $\Gamma_{\rm SN}$ & $M_{\rm Al-26}^{\rm eq}$ & $M_H$ & $\tau_{\rm gas}$ & $X (\Al26)^a$ & $\displaystyle \frac{^{26}\rm Al}{^{27}\rm Al}^b$ & $\zeta_{\rm Al-26}(e^+)^c$ & $\zeta_{\rm Al-26}(e^+ \gamma)^d$\\ 
                                      & ($\Msun\ \yr^{-1}$) & (yr$^{-1}$) & ($\Msun$) & ($\Msun$) & ($\Myr$) & & & $(\sec^{-1})$ & $(\sec^{-1})$\\
\hline
Milky Way ($z = 0$)$^e$       & 3.0   & 0.019                & 2.8   & $4.5 \times 10^9$ & 1500 & $2.4 \times 10^{-11}$ & $9.4 \times 10^{-6}$ & $1.9 \times 10^{-20}$ & $7.1 \times 10^{-20}$\\
Galactic Centre CMZ$^f$       & 0.071 & $4.6 \times 10^{-4}$ & 0.067 & $3 \times 10^7$   & 420  & $8.6 \times 10^{-11}$ & $3.4 \times 10^{-5}$ & $6.9 \times 10^{-20}$ & $2.6 \times 10^{-19}$\\
NGC 253 core$^{g,h}$          & 3.6   & 0.023                & 3.3   & $3 \times 10^7$   & 8.3  & $4.3 \times 10^{-9}$  & $1.8 \times 10^{-3}$ & $3.4 \times 10^{-18}$ & $1.3 \times 10^{-17}$\\
M82$^{g,i}$                   & 10.5  & 0.067                & 9.8   & $2 \times 10^8$   & 19   & $1.9 \times 10^{-9}$  & $7.5 \times 10^{-4}$ & $1.5 \times 10^{-18}$ & $5.7 \times 10^{-18}$\\
Arp 220 nuclei$^j$            & 50    & 0.3                  & 44    & $10^9$            & 20   & $1.7 \times 10^{-9}$  & $6.7 \times 10^{-4}$ & $1.3 \times 10^{-18}$ & $5.0 \times 10^{-18}$\\
Submillimeter galaxy$^k$      & 1000  & 6.4                  & 930   & $2.5 \times 10^{10}$ & 25   & $1.4 \times 10^{-9}$ & $5.7 \times 10^{-4}$ & $1.1 \times 10^{-18}$ & $4.3 \times 10^{-18}$\\
BzK galaxies$^l$              & 200   & 1                    & 200   & $7 \times 10^{10}$   & 400  & $1 \times 10^{-10}$  & $4 \times 10^{-5}$  & $8 \times 10^{-20}$ & $3 \times 10^{-19}$\\
\hline
\end{tabular}
\\$^a$: Mean abundance of \Al26, calculated assuming the \Al26 is well-mixed with the gas and resides there for a full decay time (instead of, for example, a wind-crossing time).
\\$^b$: Calculated assuming Solar metallicity with $\log_{10} [N(^{27}{\rm Al})/N(H)] = -5.6$.
\\$^c$: Ionization rate from \Al26 with the derived abundance, with ionization only from MeV positrons released by the decay, assuming effective stopping.
\\$^d$: Ionization rate from \Al26 with the derived abundance, where ionization from the 1.809 MeV decay line and 0.511 keV positron annihilation gamma rays is included, assuming they are all stopped.
\\$^e$: Supernova rate and gas mass from \citet{Diehl06}; SFR calculated from supernova rate using Salpeter IMF for consistency.
\\$^f$: Inner 100 pc of the Milky Way.  SFR and $\Gamma_{\rm SN}$ from IR luminosity in \citet*{Launhardt02}; gas mass from \citet{Molinari11}.  \citet{PiercePrice00} gives a gas mass of $5 \times 10^7\ \Msun$.
\\$^g$: SFR and $\Gamma_{\rm SN}$ from IR luminosity in \citet{Sanders03}.
\\$^h$: Gas mass from \citet*{Harrison99}.
\\$^i$: Gas mass from \citet{Weiss01}.
\\$^j$: Assumes IR luminosity of $3 \times 10^{11}\ \Lsun$ for SFR and $\Gamma_{\rm SN}$ and gas mass given in \citet{Downes98}.
\\$^k$: Typical gas mass and SFR of {\bfnop submillimetre} galaxies from \citet{Tacconi06}.
\\$^l$: Mean SFR and gas mass of the 6 BzK galaxies in \citet{Daddi10}, which are representative of main sequence galaxies at $z \approx 1.5$.
\end{minipage}
\end{table*}

The true starbursts, driven by galaxy mergers and galaxies interacting with each other, represent about $\sim 10\%$ of star formation at all redshifts \citep{Rodighiero11,Sargent12}.  They have SSFRs that are up to an order of magnitude higher than $z = 2$ normal galaxies.  The mean, background abundances of SLRs in starbursts are therefore about 100 times greater than the present day Milky Way.  

I show the \Al26 abundances in some nearby starburst galaxies in Table~\ref{table:Al26Abundances}.  In the Galactic Centre region, the \Al26 abundance is only twice that of the present Milky Way as a whole.  However, the \Al26 abundances are extremely high in the other starbursts, $\sim 2 \times 10^{-9}$, about twenty times that of the primordial Solar nebula.  The $^{26}{\rm Al}/^{27}{\rm Al}$ ratio in these starbursts is also very high.  Assuming Solar metallicity with an $^{27}$Al abundance of $\log_{10} [N(^{27}{\rm Al})/N(H)] = -5.6$ \citep{Diehl06}, this ratio is $\sim (0.6 - 1.8) \times 10^{-3}$.  Again, this ratio for Solar metallicity gas is $\sim 10 - 30$ times higher than that of the early Solar Nebula, $\sim 5 \times 10^{-5}$.

\section{Systematic Uncertainties}
\subsection{Effects of Variable Star-Formation Rates}

The steady-state abundance (equation~\ref{eqn:MeqSLR}) is only appropriate when the star-formation rate is slowly varying on time-scales of a few Myr.  Since young stellar populations produce SLRs for several Myr, and since \Al26 and \Fe60 themselves survive for $\ga 1\ \Myr$, the injection rate of SLRs is smoothed over those time-scales (equation~\ref{eqn:QSLR}).  Very high frequency fluctuations in the SFR therefore have little effect on the abundance of SLRs.  In the opposite extreme, when the fluctuations in SLRs are slow compared to a few Myr, we can simply take the present SFR and use it in equation~\ref{eqn:MeqSLR} for accurate results.  However, intermediate frequency variability invalidates the use of equation~\ref{eqn:MeqSLR}, and can result in the SLR abundance being out of phase with the SFR.

Normal main sequence galaxies at high redshift built up their stellar populations over Gyr times, evolving secularly \citep[c.f.,][]{Wuyts11}.  They are also large enough to contain many stellar clusters, so that stochastic effects average out.  It is reasonable to suppose that they have roughly constant SFRs over the relevant time-scales.  True starbursts, on the other hand, are violent events that last no more than $\sim 100\ \Myr$, as evinced by their short $\tau_{\rm gas}$.  They are relatively small, so stochastic fluctuations in their star-formation rates are more likely.  \citet{Forster03} studied the nearby, bright starburst M82 and concluded that its star-formation history is in fact bursty.  The star-formation histories of other starbursts are poorly known, but \citet{Mayya04} present evidence for large fluctuations on $\sim 4\ \Myr$ times.

I estimate the magnitude of these fluctuations for the prototypical starburst M82 with the full equation for SLR mass in a one-zone model (equation~\ref{eqn:MeqSLRFull}).  The solution to equation~\ref{eqn:MeqSLRFull} for $M_{\rm SLR}$ is
\begin{equation}
M_{\rm SLR} (t) = \int_{-\infty}^t {\rm SFR}(t^{\prime}) \times m_{\rm SLR}(t^{\prime}) dt^{\prime},
\end{equation}
where 
\begin{equation}
m_{\rm SLR}(t^{\prime}) = \int_{-t^{\prime}}^0 Y_{\rm SLR}(t^{\prime\prime}) \exp\left(-\frac{t^{\prime} - t^{\prime\prime}}{\tau_{\rm SLR}}\right) dt^{\prime\prime}.
\end{equation}
The quantity $m_{\rm SLR}(t^{\prime})$ represents the SLR mass in the ISM from a coeval stellar population of unit mass and age $t^{\prime}$.  It is given by \citet{Cervino00} and \citet{Voss09} for \Al26 and $^{60}$Fe.

I use the star-formation history derived by \citet{Forster03} for the `3D region' of M82, which consists of two peaks at 4.7 Myr ago and 8.9 Myr ago.  The peaks are modelled as Gaussians with the widths given in \citet{Forster03} (standard deviations $\sigma$ of 0.561 Myr for the more recent burst, and 0.867 Myr for the earlier burst).  I convert the star-formation rate from a Salpeter IMF from 1 to 100$\ \Msun$ given in \citet{Forster03} to a Salpeter IMF from 0.1 to 100$\ \Msun$ for consistency with the rest of the paper.\footnote{I ignore the relatively small difference between the upper mass limit of 100$\ \Msun$ in \citet{Forster03} and 120$\ \Msun$ in \citet{Cervino00} and \citet{Voss09}.  Since stars with masses 100 to 120$\ \Msun$ can affect stellar diagnostics, converting to that IMF may require an adjustment to the star-formation history beyond a simple mass scaling.}  This region does not include the entire starburst; it has roughly $1/3$ of the luminosity of the starburst, but the stellar mass formed within the 3D region over the past 10 Myr gives an average SFR of 10 $\Msun\ \yr^{-1}$ in the \citet{Forster03} history.  Note that \citet*{RodriguezMerino11} derives a different age distributions for stellar clusters (compare with \citealt{Satyapal97}).  \citet{Strickland09} has also argued that the star-formation history of M82's starburst core is not well constrained before 10 Myr ago (as observed from Earth), and may have extended as far back as 60 Myr ago.  Thus, I take the \citet{Forster03} history merely as a representative example of fluctuating SFRs.

\begin{figure}
\centerline{\includegraphics[width=8cm]{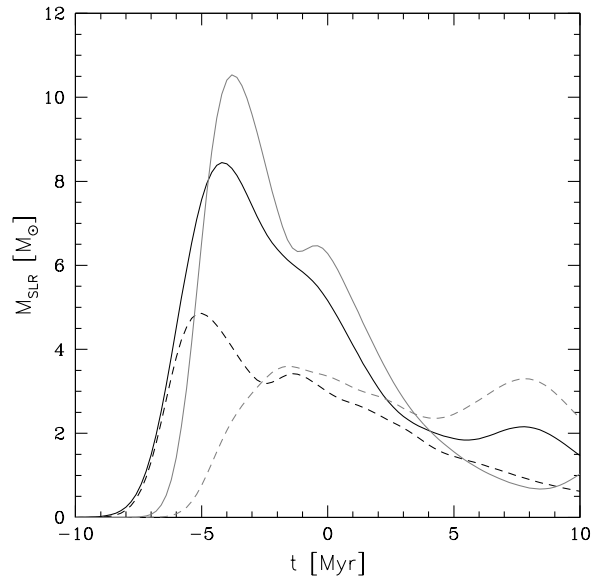}}
\caption{History of the SLR masses in M82's `3D region' ISM for the star-formation history given in \citet{Forster03}.  The black lines are for \Al26 and grey lines are for $^{60}$Fe; solid lines are using the yields in \citet{Voss09} and dashed lines are using the \citet{Cervino00} yields.  We presently observe M82 at $t = 0$; I assume there are no bursts of star-formation after then, so that the masses inevitably decay away. \label{fig:M82SLRHistory}}
\end{figure}

The calculated \Al26 (black) and \Fe60 (grey) masses are plotted in Fig.~\ref{fig:M82SLRHistory}.  At first, there is no SLR mass in the starburst, because it takes a few Myr for SLR injection to start.  With the \citet{Voss09} yields, the SLR masses rise quickly and peak $\sim 5\ \Myr$ ago (as observed from Earth).  The SLR masses drop afterwards.  Yet they are still within a factor of 1.7 of their peak values even now, $\sim 5\ \Myr$ after the last star-formation burst.  If there is no further star-formation, the SLRs will mostly vanish over the next 10 Myr.  The \citet{Cervino00} yields predict a greater role for supernovae from lower mass stars, so the fluctuations are not as great; the \Fe60 mass remains roughly the same even 10 Myr from now.  As long as there has been recent star-formation in the past $\sim 5\ \Myr$, the SLR abundances are at least half of those predicted by the steady-state assumption.

\begin{figure}
\centerline{\includegraphics[width=8cm]{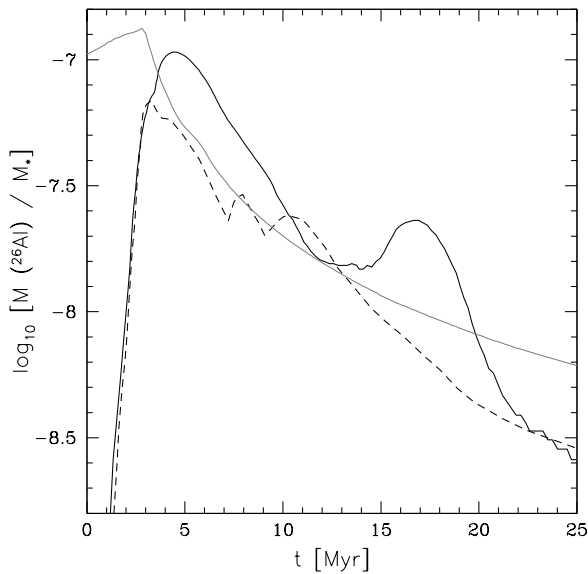}}
\caption{How the bolometric luminosity traces \Al26 mass for a coeval stellar population with age $t$.  The grey line is the predicted steady state \Al26 mass I would predict from the bolometric luminosity of the population, whereas the black lines are the actual mass of \Al26 (solid for \citealt{Voss09} and dashed for \citealt{Cervino00}).\label{fig:LBolVsMAl26}}
\end{figure}

There is a more fundamental reason to expect that the steady-state SLR abundances are roughly correct for starbursts.  A common way of estimating star-formation rates in starbursts is to use the total infrared luminosity \citep{Kennicutt98}, which is nearly the bolometric luminosity for these dust-obscured galaxies.  Young stellar populations, containing very massive stars, are brighter and contribute disproportionately to the bolometric luminosity.  Therefore, both the luminosity and the SLR abundances primarily trace young stars.  To compare the bolometric luminosity, I ran a Starburst99 (v6.04) model of a $Z = 0.02$ metallicity coeval stellar population with a Salpeter IMF ($dN/dM \propto M^{-2.35}$) between 0.1 and 120$\ \Msun$ \citet{Leitherer99}.  I then calculate the SFR that would be derived from these {\bfnop luminosities} using the \citet{Kennicutt98} conversion, and then from that, the expected steady-state SLR masses from equation~\ref{eqn:MeqSLR}.  The `bolometric' \Al26 masses are compared to the actual masses in Fig.~\ref{fig:LBolVsMAl26}.

Although the very youngest stellar populations are bright but not yet making SLRs, the bolometric luminosity (grey) is a good tracer of \Al26 mass (black) for stellar populations with ages between 3 and 20 Myr.  For most of the interval, the bolometric \Al26 masses are within a factor 2 of the actual masses.  For populations between 15 Myr and 20 Myr, the \citet{Voss09} and \citet{Cervino00} predictions envelop the bolometric \Al26 masses.  At 20 to 25 Myr old, the bolometric \Al26 masses are about twice the true masses.  For older populations still, the true \Al26 masses finally die away while the bolometric luminosity only slowly declines.  Note that, if stars have been forming continuously for the past 100 Myr, over half of the luminosity comes from stars younger than 20 Myr.  Thus, the use of the bolometric luminosities introduces a factor $\la 3$ systematic error.

In short, the use of bolometric luminosity as a SFR indicator, and the natural variability in the star-formation rates of starbursts can lead to overestimations of the SLR abundances by a factor $\sim 3$.  But I estimate the SLR abundances of true starbursts are a hundred times higher than in the present Milky Way (equation~\ref{eqn:XAl26Numer} and Table~\ref{table:Al26Abundances}).  The ratio is so great that the systematic effects do not undermine the basic conclusion that SLR abundances are much larger in true starbursts.

\subsection{Are SLRs mixed quickly enough into the gas?}
\label{sec:Mixing}
Although the average levels of SLRs in starbursts and high-$z$ normal galaxies are high, that does not by itself mean the SLRs influence the environments for star-formation.  While SLRs can play an important role in star-forming regions, by elevating the ionization rates and by being incorporated into solid bodies, SLRs trapped in ionized gas are irrelevant for these processes.  

The mixing of metals from young stars into the ISM gas mass is usually thought to be very slow in the present Milky Way, compared to SLR lifetimes.  The massive stars responsible for making SLRs often live in star clusters, which blow hot and rarefied bubbles in the ISM.  Supernovae also excavate the coronal phase of the ISM \citep{McKee77}.  Turbulence within the bubbles mixes the SLRs and homogenizes their abundances \citep[e.g.,][]{Martin10}, over a time scale $t_{\rm mix} \approx L / v_{\rm turb}$, where $L$ is the outer scale of turbulence (typical size of the largest eddies) and $v_{\rm turb}$ is the turbulent speed \citep{Roy95,Pan10}.  The large outer scale of turbulence, $\sim$100 -- 1000$\ \pc$, and the slow turbulent speeds ($\sim 5$--$10\ \kms$) in the Milky Way imply mixing times of $\sim 10$ -- $200\ \Myr$.  Even if the \Al26 is homogenized within the superbubbles, this low density hot gas requires a long time to mechanically affect cold star-forming clouds, because of the large density contrast \citep{deAvillez02}.  Mixing between the phases, particularly warm and hot gas, is accelerated by Rayleigh-Taylor and Kelvin-Helmholtz instabilities \citep{Roy95}, but overall, mixing takes tens of Myr to operate in the Milky Way (\citealt{deAvillez02}; see also \citealt{Clayton83}, where mixing times are between the warm ISM from evaporated H I clouds, cool and large H I clouds, and molecular clouds).  Thus, SLRs like \Al26 are thought to decay long before they are mixed thoroughly with the star-forming gas.  Indeed, studies of the abundances of longer lived isotopes in the primordial Solar system supports longer mixing times of $\sim 50$ -- 100$\ \Myr$ \citep[e.g.,][]{Meyer00,Huss09}.  

It is thought that these obstacles existed, at least qualitatively, in the $z \approx 0.45$ Milky Way, when the Solar system formed.  These problems are part of the motivation for invoking a local source of SLRs, including energetic particles from the Sun itself \citep{Lee98}, injection from a nearby AGB star (\citealt*{Busso03}), or injection from an anomalously nearby supernova \citep{Cameron77} or Wolf-Rayet star (\citealt*{Arnould97}).  Recently, though, several authors proposed models that might overcome the mixing obstacle, where young stars are able to inject SLRs into star-forming clouds.  A motivation behind these models is the idea that molecular clouds are actually intermittent high density turbulent fluctuations in the ISM \citep[e.g.,][]{MacLow04}, and the supernovae that partly drive the turbulence -- indirectly forming the molecular clouds -- also are the sources of SLRs \citep{Gounelle09}.   In the model of \citet{Gounelle09}, old superbubbles surrounding stellar clusters form into molecular clouds after {\bfnop ploughing} through the ISM for $\sim$10 Myr.  Supernovae continue going off in the star clusters, adding their SLRs into these newborn molecular clouds \citep[see also][]{Gounelle12}.  Simulations by \citet*{Vasileiadis13} also demonstrate that SLRs from supernovae going off very near giant molecular clouds are mixed thoroughly with the star-forming gas.  On a different note, \citet{Pan12} argued that supernovae remnants are clumpy, and that clumps could penetrate into molecular clouds surround star clusters and inject SLRs.  

If these scenarios are common, then a large fraction of the produced SLRs reaches the star-forming gas before decaying.  In fact, these mechanisms may be so efficient that SLRs are concentrated only into star-forming molecular gas, a minority of the Milky Way gas mass.  If so, then the abundance of SLRs within Galactic molecular gas ($M_{\rm SLR} / M_{\rm H2}$) is greater than the mean background level ($M_{\rm SLR} / M_{\rm gas}$); in this way, SLR abundances could reach the elevated levels that existed in the early Solar system \citep{Gounelle09,Vasileiadis13}.

{\bfnop On the other hand, young stars can trigger star-formation in nearby gas without polluting them.  This can occur when a shock from a supernova or from an overpressured H II region propagates into a molecular cloud, causing the cores within it to collapse (\citealt{Bertoldi89}; \citealt*{Elmegreen95}).  This process has been inferred to happen in the Upper Scorpius OB association \citep{Preibisch99}.  Since the cores are pre-existing, they may not be enriched with SLRs (although supernova shocks can also inject SLRs into a molecular cloud; see \citealt{Boss10} and \citealt{Boss12}).  The triggering can also occur before any supernovae enrich the material with SLRs.  Star formation can also be triggered when a shock from a H II region sweeps up a shell of material, which eventually becomes gravitationally unstable and collapses (\citealt{Elmegreen77}; see also \citealt{Gritschneder09}).}

I note, however, that the homogeneity of the \Al26 abundance in the early Solar system is controversial; if the abundance was inhomogeneous, that is inconsistent with efficient SLR mixing within the Solar system's progenitor molecular cloud.  Although \citet*{Villeneuve09} conclude that \Al26 had a constant abundance, \citet{Makide11} find that the \Al26 abundance varied during the earliest stages of Solar system formation, when {\bfnop aluminium} first condensed from the Solar nebula.  

What is even less clear, though, is how similar the Milky Way is to starbursts and the massive high-$z$ normal galaxies that host the majority of the cosmic star formation.  As in the Milky Way, supernovae in starbursts like M82 probably blast out a hot phase.  But the hot phase escapes in a rapid hot wind in starbursts with high star-formation densities ($\ga 0.1\ \Msun\ \yr^{-1}\ \kpc^{-2}$; \citealt{Chevalier85,Heckman90}).  There is direct evidence for this hot wind from X-ray observations \citep{Strickland07,Strickland09}.  Furthermore, supernova remnants are observed to expand quickly in M82, implying that they are going off in a very low density environment (e.g., \citealt{Beswick06}; compare with \citealt{Chevalier01}).  Cool and warm gas is observed outflowing alongside the hotter wind, possibly entrained by the hot wind \citep{Strickland09}.  Whereas the edges of superbubbles eventually cool and fade back into the ISM in the Milky Way after a 10 -- 100 Myr, any superbubble gas that does cool in these starbursts could still be pushed out by the wind.  If the SLRs are trapped in the wind, they reside in these starbursts for only a few hundred kyr.  

But the ISM conditions in the starbursts are much different, with higher densities in all phases, higher pressures, and higher turbulent speeds \citep[e.g.,][]{Smith06}.  Starbursts are physically small, with radii of $\sim 200\ \pc$ at $z = 0$ -- comparable to the size of some individual superbubbles in the Milky Way.  The eddy sizes therefore must be smaller and mixing processes could be faster.  To demonstrate just how small superbubbles are in starbursts, \citet*{Silich07} modelled the superbubble surrounding the star cluster M82 A-1 in M82, which has a mass of $2 \times 10^6\ \Msun$.  They find that the wind propagates only for a few parsecs before being shocked and cooled.  Stellar ejecta in the core of the cluster also cool rapidly in their model.  The turbulent mixing time is therefore much smaller, $\sim 200\ \kyr$ for an eddy length of 10 pc and a turbulent speed of $50\ \kms$.

Conditions are even more extreme in present day compact Ultraluminous Infrared Galaxy starbursts, where the ISM is so dense ($\sim 10^4\ \cm^{-3}$; \citealt{Downes98}) that a hot phase may be unable to form (\citealt{Thornton98}; \citealt*{Thompson05}).  Instead, the ISM is almost entirely molecular \citep{Downes98}.  Indeed, observations of supernovae in Arp 220 indicate they are going off in average density molecular gas \citep{Batejat11}.  Supernovae remnants fade within a few tens of kyr into the ISM, due to powerful radiative losses \citep{McKee77}.  The SLRs then are incorporated into the molecular ISM in a turbulent mixing time, the whole process taking just a few hundred kyr.  The main uncertainty is then, not whether the SLRs are injected into the molecular gas, but whether these SLR-polluted regions of the molecular gas fill the entire starburst.  Turbulent mixing smooths abundances over regions the size of the largest eddies \citep{Pan10}, but if the distribution of SLR injection sites varies over larger scales, the final SLR abundance may also vary on these large scales.

We know very little about the conditions in high redshift galaxies.  Star formation in main sequence galaxies is dominated by massive galaxies with large star-formation rates at high redshift.  These massive galaxies are several kpc in radius, but contain large amounts of molecular gas \citep{Daddi10}.  They also have large star-formation densities and host winds.  In the more extreme galaxies, radiative bremsstrahlung losses stall any hot wind \citep{Silich10}.  Turbulent speeds in these galaxies are high \citep{Green10}, implying faster turbulent mixing than in the Milky Way.  But it is not clear which phase the SLRs are injected into or how long it takes for them to mix throughout star formation regions.  The effects of clustering in the ISM is also uncertain, but it probably is important in these galaxies, where huge clumps ($\ga 10^8\ \Msun$ and a kpc wide) are observed \citep[e.g.,][]{Genzel11}.

To summarize, while there are reasons to expect that most of the SLRs synthesized by young stars in the Milky Way decay before reaching star-forming gas, this is not necessarily true in starbursts or high-$z$ normal galaxies.  Turbulent mixing is probably fast, at least in compact starbursts which are physically small.  On the other hand, winds might blow out SLRs before they reach the star-forming gas, at least in the weaker starbursts.  Clearly, this issue deserves further study.

\section{Implications}
\subsection{Implications for the early Solar system and Galactic stars of similar age} 
The rapid evolution in SSFRs implies that Galactic background SLR abundances were up to twice as high during the epoch of Solar system formation (4.56 Gyr ago; $z \approx 0.44$).  If the evolution of the Milky Way's gas fraction was comparable to that in observed massive main sequence galaxies, the enhancement may have been only $\sim 50\%$ above present values (see the discussion in section~\ref{sec:MSGalaxies}; \citealt{Genzel10}).  The inferred primordial abundances of $^{60}$Fe in the Solar system are in fact up to twice as high as in the contemporary Milky Way, as determined with gamma-ray lines \citep{Williams08,Gounelle08}.  This overabundance is cited as evidence for an individual, rare event enriching the early Solar system, or the gas that formed into the Solar system, with SLRs.  However, my calculations show this is not necessarily the case: the twice high abundances of $^{60}$Fe in the early Solar system can arise \emph{simply because the Galaxy was more efficient at converting gas into stars 4.5 Gyr ago.} 

The primordial abundance of \Al26 in the Solar system was about six times higher than the mean Galactic value at present \citep{Diehl06}, or three times higher than the mean Galactic abundance at $z = 0.44$ assuming that equation~\ref{eqn:fGas} holds.  Even so, the normal effects of galaxy evolution are a potential contributor to the greater \Al26 abundances, assuming efficient mixing of \Al26 with the molecular gas of the Milky Way.  Furthermore, the high abundances of \Al26 in the early Solar system are actually typical of star-formation at $z \approx 1 - 2$ -- when most cosmic star-formation occurred.  In this sense, the early Solar system's \Al26 abundance may be normal for most planetary systems in the Universe.  

As I have discussed in Section~\ref{sec:Mixing}, it is not clear whether the background abundances of \Al26 and other SLRs actually represent those of typical star-forming gas; if mixing takes more than a few Myr, these SLRs could not have affected star and planet formation \citep{Meyer00,deAvillez02,Huss09}.  But although there may have been a wide distribution of abundances if mixing is inefficient, the mean of the distribution is still higher simply because there were more supernovae and young stars per unit gas mass.  Thus, a greater fraction of star formation occurred above any given threshold in the past.  In addition, the Galactic background level can be meaningful if a large fraction of the SLRs from young stars make it into the cold gas, and \citet{Gounelle09}{\bfnop,} \citet{Gounelle12}, \citet{Pan12}, and \citet{Vasileiadis13} have presented mechanisms where this can happen.  Although there is suggestive evidence that these mechanisms did not operate for the Solar system \citep{Makide11}, there is no reason they could not have worked for other star systems of similar ages.  Then the Solar system's relatively high abundance of SLRs, and \Fe60 in particular, may be common for Galactic stars of similar ages, even if through a different process.

Finally, as I noted, these conclusions depend on how the yields of SLRs from massive stars change with metallicity, and what the mean Galactic metallicity was at the epoch of Solar system formation.

\subsection{The Ionization Rate and Physical Conditions in Starbursts' Dense Clouds}
Radioactive decay from SLRs injects energy in the form of daughter particles into the ISM.  The decay particles, with typical energies of order an MeV, ionize gas and ultimately heat the ISM, if they do not escape entirely.  The high abundances of SLRs, including \Al26, can alter the ionization state of molecular gas in these galaxies.  The ionization rate, in particular, is important in determining whether magnetic fields can thread through the densest gas.  I focus here on the contribution from \Al26, which dominated the ionization rate from SLRs in the primordial Solar system \citep{Umebayashi09}.  For the sake of discussion, I assume that the SLRs are well-mixed into the gas, despite the uncertainties (section~\ref{sec:Mixing}).

Each \Al26 decay releases an energy $E_{\rm decay}$ into the ISM in the form of MeV positrons and gamma rays.  If each atom in the ISM takes an energy $E_{\rm ion}$ to be ionized, each \Al26 decay can therefore ionize $E_{\rm decay} / E_{\rm ion}$ atoms.  In 82\% of the decays, a positron with kinetic energy of 1.16 MeV is released, and the positron is slowed by ionization losses \citep{Finocchi97}.  The minimum energy per decay, after accounting for this branching ratio, that goes into ionization is $E_e^{\rm min} = 0.95\ \MeV$, when the medium stops the positron (inflight annihilation losses are negligible at these energies; \citealt{Beacom06}).  The annihilation of the positron with {\bfnop an} electron in the ISM produces gamma rays of total energy $2 \times 0.511 = 1.022$ MeV.  In addition, in very nearly all \Al26 decays, a 1.809 MeV gamma ray is produced.  These gamma rays only interact with the ISM over columns of several $\gcm2$, so only in particularly dense regions will they contribute to the ionization \citep{Finocchi97}.  When they do, $E_{\rm decay}^{e \gamma} = 3.60\ \MeV$.  \citet{Stepinski92} gives $E_{\rm ion}$ as 36.3 eV, so that the ionization rate is
\begin{equation}
\zeta_{\rm Al-26} = \frac{X_{\rm Al-26} E_{\rm decay}}{(36.3\ \eV) \tau_{\rm decay}}.
\end{equation}

In terms of gas consumption time, the ionization rate from \Al26 is
\begin{equation}
\zeta = (1.4 - 5.1) \times 10^{-18}\ \sec^{-1}\ \left(\frac{\tau_{\rm gas}}{20\ \Myr}\right)^{-1}.
\end{equation}
My results for the mean ionization rate from \Al26 of some characteristic starbursts are shown in Table~\ref{table:Al26Abundances}; they are in the range $10^{-18} - 10^{-17}\ \sec^{-1}$.  The maximal ionization rates are roughly an order of magnitude higher than that found in early Solar system, a dense environment with $\zeta \approx (0.6 - 1) \times 10^{-18}\ \sec^{-1}$ \citep{Finocchi97,Umebayashi09}.  

Even if the \Al26 is in the cold star-forming gas, it could actually be condensed into dust grains instead of existing in the gas phase.  Yet the decay products still escape into the ISM from within the grain.  The attenuation of gamma rays at $\sim 1$ MeV is dominated by Compton scattering, requiring columns of a few $\gcm2$ to be important.  Thus, gamma rays pass freely through dust grains that are smaller than a {\bfnop centimetre}.  The energy loss rate of relativistic electrons or positrons in neutral matter is approximately
\begin{equation}
\frac{dK}{ds} = \frac{9}{4} m_e c^2 \sigma_T \sum_j \frac{\rho_j Z_j}{A_j m_H} \left[\ln \frac{K + m_e c^2}{m_e c^2} + \frac{2}{3} \ln \frac{m_e c^2}{\mean{E_j}}\right]
\end{equation}
from \citet{Schlickeiser02}, where $K$ is the particle kinetic energy, $s$ is the physical length, and $\sigma_T$ is the Thomson cross section.  The sum is over elements $j$; for heavy elements $Z_j \approx A_j / 2$, $\rho_j$ is the partial density of each element within the grain, and $\mean{E_j}$ is related to the atomic properties of the element.  I take the bracketed term to have a value $\sim 5$ and $\rho \approx 3\ \gcm2$.  Then the stopping length is $K / (dK / ds) \approx 0.3\ \cm\ (K / \MeV)$, much bigger than the typical grain radius.  Thus, \Al26 (and other SLRs) in dust grains still contribute to the ionization of the ISM.

On the other hand, are the positrons actually stopped in starburst molecular clouds, or do they escape?  For neutral interstellar gas, the stopping column of MeV positrons is $\sim K \rho / (dK / ds) \approx 0.2\ \gcm2$ {\bfnop through ionization and excitation of atoms.}  \citep{Schlickeiser02}.  {\bfnop In cold molecular gas, ionization and excitation continue to cool the positrons until $K \approx 10\ \eV$, at which point they start annihilating by charge exchange reactions or they thermalize \citep*{Guessoum05}.}  The column densities of starbursts range from $\sim 0.1 - 10\ \gcm2$ \citep[e.g.,][]{Kennicutt98}, and the columns through individual molecular clouds are expected to be similar to those of the galaxies (\citealt*{Hopkins12}).  In the denser molecular clouds, positrons are stopped even if they are not confined at all.  In massive main sequence galaxies, the columns are $\sim 0.1\ \gcm2$ \citep{Daddi10}, insufficient to stop positrons moving in straight lines.  If magnetic waves with a wavelength near the positron gyroradius scale exist in these clouds, they efficiently scatter positrons and confine them.  {\bfnop As a result, the propagation of the positrons can be described with a diffusion constant, as widely used when interpreting the Galactic GeV positron spectrum (e.g., \citealt{Moskalenko98}; \citealt*{Hooper09}), although it is unclear how relevant these studies are for MeV positrons \citep{Jean09,Prantzos11,Martin12}.} However, these waves are probably damped quickly in dense neutral gas (see {\bfnop \citealt{Jean09}; \citealt*{Higdon09};} \citealt{Prantzos11} and references therein).  On the other hand, positrons move along magnetic field lines, and if the lines themselves are twisted on a scale $\lambda_B$, the positrons are forced to random walk with a similar mean free path {\bfnop \citep{Martin12}}.  As long as $\lambda_B$ is less than about a third of the molecular cloud size, then positrons are stopped in these galaxies.

If it is well mixed with the molecular gas, does \Al26 dominate the ionization rate in molecular gas in starbursts, and if so, what physical conditions does it induce?  The starburst \Al26 ionization rates are about an order of magnitude lower than the canonical cosmic ray-sustained ionization rates in most Milky Way molecular clouds, but in some of the densest Galactic starless cores, the ionization rate drops to $\sim 10^{-18}\ \sec^{-1}$ \citep{Caselli02,Bergin07}.  Cosmic rays in starbursts can sustain much higher ionization rates (up to $\sim 10^{-14}\ \sec^{-1}$; c.f., \citealt{Papadopoulos10-CRDRs}), but cosmic rays can be deflected by magnetic fields, possibly preventing them from penetrating high columns.  {\bfnop Aside from cosmic rays themselves, starbursts are also bright sources of GeV gamma rays, which are generated when cosmic rays interact with the prevalent gas \citep[e.g.,][]{Ackermann12}.  These gamma rays can penetrate molecular clouds and induce low levels of ionization \citep{Lacki12-GRDRs}.}  In \citet{Lacki12-GRDRs}, I found that the gamma-ray ionization rate in starbursts can be anywhere from $10^{-22} - 10^{-16}\ \sec^{-1}$, with values of $\sim (1 - 3) \times 10^{-19}\ \sec^{-1}$ in M82 and $\sim (5 - 8) \times 10^{-17}\ \sec^{-1}$ in Arp 220's radio nuclei.  In the dense clouds of most starbursts, \Al26 radioactivity could exceed the ionization rate over gamma rays, setting a floor to the ionization rate.  In the most extreme starbursts, with mean gas surface densities of $\ga 3\ \gcm2$ (c.f. equation 10 of \citealt{Lacki12-GRDRs}), however, gamma-ray ionization is more important, since the gamma-ray ionization rate depends strongly on the density of gas and compactness of the starbursts.  Unlike the uncertainty of how SLRs and their positron decay products are transported and mixed with the gas of starbursts, gamma rays propagate relatively simply, so the gamma ray ionization rates are more secure.

An \Al26-dominated ionization rate has implications for the physical conditions of star-forming clouds.  According to \citet{McKee89}, the ionization fraction of a cloud with hydrogen number density $n_H$ is
\begin{equation}
\label{eqn:xELow}
x_e \approx 1.4 \times 10^{-8} \left(\frac{\zeta}{10^{-18}\ \sec^{-1}}\right)^{1/2} \left(\frac{n_H}{10^4\ \cm^{-3}}\right)^{-1/2}
\end{equation}
when the ionization rate is low.  We see that the ionization fraction of cloud with density $n_H = 10^4\ \cm^{-3}$ {\bfnop is} $(1 - 4) \times 10^{-8}$, if the ionization is powered solely by \Al26 decay.  For these ionization fractions, the ambipolar diffusion time of a molecular core, the time for magnetic fields to slip from the gas, is a few times its free-fall time.  Since clouds with strong magnetic fields do not collapse until the field slips away by ambipolar diffusion \citep{Mestel56,Mouschovias76}, this means that \Al26-ionized clouds in starbursts collapse quasi-statically, as in the Milky Way.  

On the other hand, the energy injection from \Al26 has essentially no effect on the gas temperature. \citet{Papadopoulos10-CRDRs} gives the minimum gas temperature of gas as:
\begin{equation}
T_k^{\rm min} = 6.3\ \Kelv\ [(0.0707 n_4^{1/2} \zeta_{-18} + 0.186^2 n_4^3)^{1/2} - 0.186 n_4^{3/2}]^{2/3},
\end{equation}
which was derived under the assumption that there is no heating from interactions with dust grains or the dissipation of turbulence in the gas, for gas with density $n_4 = (n_H / 10^4\ \cm^{-3})$ and ionization rate $\zeta_{-18} = (\zeta / 10^{-18}\ \sec^{-1})$.  In typical starbursts, I find that \Al26 decay alone heats gas of density $n_H = 10^{4}\ \cm^{-3}$ to $\sim 2 - 5\ \Kelv$ ($0.1 - 0.5\ \Kelv$ for $n_H = 10^6\ \cm^{-3}$).  As I note in \citet{Lacki12-GRDRs}, under such conditions, dust heating is more likely to set the temperature of the gas than ionization, raising the temperature to the dust temperature for densities $\ga 40000\ \cm^{-3}$.

\section{Conclusions}

The high SSFRs of starbursts and high-$z$ normal galaxies implies high abundances of \Al26 and other SLRs in their ISMs.  In true starbursts, these abundances are enormous, with $X (\Al26) \approx 10^{-9}$ and $^{26}$Al/$^{27}$Al $\approx 10^{-3}$.  The SSFRs of normal galaxies evolve rapidly with $z$; even taking into account higher gas fractions, the SLR abundances were about 3 -- 10 times higher at $z \approx 2$ than in the present Milky Way.  Even at the epoch of Solar system formation, the mean SLR abundance of the Milky Way was 1.5 to 2 times as high as at the present (Fig.~\ref{fig:NormGalaxy}).  This alone could explain the high abundances of $^{60}$Fe in the early Solar system, and reduce the discrepancy in the \Al26 abundances from a factor of $\sim 6$ to $\sim 3$.  In this way, the cosmic evolution of star-forming galaxies may have direct implications for the early geological history of the Solar system.  The first main uncertainty is whether the SLRs produced by massive stars is well-mixed with the molecular gas: they may instead be ejected by the galactic winds known to be present in starbursts and high-$z$ galaxies, or decay before they can propagate far from its injection site, or before it penetrates cold gas.  I discussed these uncertainties in section~\ref{sec:Mixing}.  The other uncertainty is how SLR yields depend on metallicity.  

The most direct way to test the high \Al26 abundances of starburst galaxies is to detect the 1.809 MeV gamma-ray line produced by the decay of \Al26, directly informing us of the equilibrium mass of \Al26.  Whether most of the \Al26 is ejected by the superwind can be resolved with spectral information.  The turbulent velocities of molecular gas in starbursts, $\sim 100\ \kms$ \citep{Downes98}, is much smaller than the bulk speed of the superwind, which is hundreds or even thousands of $\kms$ \citep{Chevalier85}.  Unfortunately, the \Al26 line fluxes predicted for even the nearest external starbursts ($\sim 10^{-8}\ \cm^{-2}\ \sec^{-1}$) are too low to detect with any planned instrument (\citealt*{Lacki12-MeV}).  However, \citet{Crocker11-Wild} have argued that the inner 100 pc of the Galactic Centre are an {\bfnop analogue} of starburst galaxies, launching a strong superwind.  The \Al26 line from this region should have a flux of $\sim 2 \times 10^{-5}\ \cm^{-2}\ \sec^{-1}$, easily achievable with possible future MeV telescopes like Advanced Compton Telescope (ACT; \citealt{Boggs06}) or Gamma-Ray Burst Investigation via Polarimetry and Spectroscopy (GRIPS; \citealt{Greiner11}).  A search for the \Al26 signal from this region may inform us on its propagation, since it is nearby and resolved spatially{\bfnop.}\footnote{\citet{Naya96} reported that the \Al26 signal from the inner Galaxy had line widths corresponding to speeds of several hundred $\kms$, but this was not verified by later observations {\bfnop by RHESSI} \citep{Smith03} {\bfnop and INTEGRAL} \citep{Diehl06-Linewidth}.  {\bfnop Instead, recent observations indicate that Galactic \Al26 is swept up in superbubbles expanding at 200 $\kms$ into the interarm regions \citep{Kretschmer13}.} However, this signal is from the entire inner Galactic disc on kiloparsec scales; the inner 100 pc of the Galactic Centre is a much smaller region and just a small part of this signal, so the kinematics of its \Al26 are currently unconstrained.  Current observations of the \Al26 decay signal from the Galactic Centre region are summarized in \citet{Wang09}.}  If the \Al26 generated in the Centre region is actually advected away by the wind, the `missing' \Al26 should be visible several hundred pc above and below the Galactic Plane near the Galactic Centre{\bfnop.}\footnote{This will also be true in other starbursts, but these starbursts would not be resolved spatially by proposed MeV telescopes.  The total amount of \Al26 line emission from other starbursts would therefore not inform us of whether it is in the starburst proper or in the superwind; other information, such as the Doppler width of the line, is necessary to determine that.}  Resolved measurements of the Galactic Centre can also inform us on whether the \Al26 is present in all molecular clouds in the region (and therefore is well-mixed), or just if it is trapped near a few injection sites.

If the SLRs do mix with the star-forming molecular gas of these galaxies, there are implications for both their star formation and planet formation.  Ionization rates of $10^{-18} - 10^{-17}\ \sec^{-1}$, like those in some Milky Way starless cores, result from the energy injection of \Al26 decay in starbursts.  While cosmic ray ionization rates can easily exceed those ionization rates by orders of magnitude in gas they penetrate into, and while gamma rays produce higher ionization rates in the most extreme starbursts like Arp 220, \Al26 might dominate the ionization rate in the dense clouds shielded from cosmic rays in typical starbursts.  In starbursts' protoplanetary discs, \Al26 can provide moderate ionization through all columns, possibly eliminating the `dead zones' where there is little accretion (e.g., \citealt{Gammie96}; \citealt*{Fatuzzo06}).  Any planetesimals that do form in starbursts may have much higher radiogenic heating from \Al26.  Admittedly, studying the geological history of planets, much less planetesimals, in other galaxies is very difficult for the forseeable future.  However, at $z \approx 2$ the Milky Way likely had background SLR abundances $\sim 10$ times higher than at present, so the effects of elevated SLR abundances may be studied in planetary systems around old Galactic stars.  

On a final point, \citet{Gilmour09} propose that the elevated abundances of \Al26 in the early Solar system are mandated by anthropic selection, since high SLR abundances are necessary for planetesimal differentiation and the loss of volatiles, but that explanation may be difficult to maintain.  If high \Al26 abundances, far from being very rare, are actually typical of high-$z$ and starburst solar systems (and indeed, much of the star-formation in the Universe's history), the anthropic principle would imply that most technological civilizations would develop in solar systems formed in these environments.  Instead of asking why we find ourselves in a system with an \Al26 abundance just right to power differentiation and evaporate volatiles, we must ask why we find ourselves in one of the rare solar systems with sufficient \Al26 that formed in a normal spiral galaxy at $z \approx 0.4$, instead of the common \Al26-enriched solar systems formed at $z \approx 2$ or in starbursts.

\section*{Acknowledgements}
During this work, I was supported by a Jansky Fellowship from the National Radio Astronomy Observatory.  The National Radio Astronomy Observatory is operated by Associated Universities, Inc., under cooperative agreement with the National Science Foundation.

\end{document}